\documentclass[prx,twocolumn,english,superscriptaddress,floatfix,longbibliography, nofootinbib]{revtex4-2}

\date{\today}
\usepackage[linesnumbered]{algorithm2e}
\usepackage[final]{hyperref}
\usepackage{graphicx, amsmath, amssymb, cleveref, xcolor, amsthm, stmaryrd, thmtools}

\usepackage{siunitx}

\crefname{defn}{Definition}{Definitions}
\crefname{lemma}{Lemma}{Lemmas}
\crefname{theorem}{Theorem}{Theorems}
\crefname{fact}{Fact}{Facts}
\crefname{corollary}{Corollary}{Corollaries}
\crefname{figure}{Figure}{Figures}

\newtheorem{theorem}{Theorem}
\newtheorem{corollary}[theorem]{Corollary}
\newtheorem{lemma}[theorem]{Lemma}
\newtheorem{defn}[theorem]{Definition}
\newtheorem{fact}[theorem]{Fact}

\DeclareMathOperator*{\argmax}{arg max}
\DeclareMathOperator{\supp}{supp}

\newcommand{\slack}{\mathrm{slack}}

\newcommand{\F}{\mathbb{F}_2}
\renewcommand{\O}{\mathcal{O}_\sigma}
\newcommand{\C}{\mathcal{C}}
\newcommand{\VNTS}{V_{\sigma}}

\begin{document}

\title{Efficient soft-output decoders for the surface code}

\author{Nadine Meister}
\affiliation{Department of Physics, Harvard University, Cambridge, MA 02138}

\author{Christopher A. Pattison}
\affiliation{Institute for Quantum Information and Matter, California Institute of Technology, Pasadena, CA 91125}

\author{John Preskill}
\affiliation{Institute for Quantum Information and Matter, California Institute of Technology, Pasadena, CA 91125}
\affiliation{AWS Center for Quantum Computing, Pasadena CA 91125}

\begin{abstract}
   Decoders that provide an estimate of the probability of a logical failure conditioned on the error syndrome (``soft-output decoders'') can reduce the overhead cost of fault-tolerant quantum memory and computation.
    In this work, we construct efficient soft-output decoders for the surface code derived from the Minimum-Weight Perfect Matching and Union-Find decoders. We show that soft-output decoding can improve the performance of a ``hierarchical code,'' a concatenated scheme in which the inner code is the surface code, and the outer code is a high-rate quantum low-density parity-check code. Alternatively, the soft-output decoding can improve the reliability of fault-tolerant circuit sampling by flagging those runs that should be discarded because the probability of a logical error is intolerably large.
\end{abstract}

\maketitle

\section{Introduction}\label{sec:Intro}

Quantum error correction will be essential for realizing large-scale quantum computers that can solve very hard problems. The currently preferred quantum error-correcting code is the surface code, which has two distinct advantages: it requires only geometrically local quantum processing in a two-dimensional layout, and it has a relatively high error threshold. However, for realistic noise, the overhead cost of fault-tolerant quantum computing with the surface code is quite daunting. 

High-rate quantum low-density parity-check (qLDPC) codes are known which are much more efficient than the surface code, but these require nonlocal processing in two dimensions. Recently, ``hierarchical codes'' were constructed \cite{pattison2023hierarchical}, in which an inner surface code is concatenated with an outer qLDPC code. In this scheme, surface code inner blocks can be swapped fault-tolerantly, enabling nonlocal syndrome extraction for the outer code, thus reducing the asymptotic overhead cost of fault tolerance. However, the task of decoding such code families is relatively unexplored.

A naive decoder for concatenated codes executes first a decoder for the inner code and then a decoder for the outer code in a black-box manner, but this approach is suboptimal. A better method is to decode both levels \emph{jointly} without discarding information \cite{poulin2006optimal}. One can apply a \textit{soft-output decoder} to each inner code block, which infers from the syndrome not only a recovery operation that returns the block to the code space, but also an estimate of the probability that the block has suffered a logical error. This soft information can then be exploited to improve the performance of the decoder applied to the outer code. 

Such a soft output is generated naturally by tensor network decoders, but unfortunately these have an exponential time complexity in general. It is thus of considerable interest to have a soft output for \emph{efficient} decoders.
A soft-output modification of the Minimum-Weight Perfect Matching (MWPM) decoder for the surface code, known as the complementary gap method, has been formulated previously \cite{hutter2014efficient,bombin2024fault,gidney2023yoked}, in which MWPM is performed using a modified syndrome graph in order to compare the weight of corrections for the two possible homology classes.
Here we offer an alternative approach, noting that the decoder organizes the error syndrome into clusters, and that the soft output can be extracted from the geometry of these clusters. 
This approach has two advantages: (1) The method is suitable for both the MWPM decoder and the Union-Find Decoder (UFD). (2) Modern MWPM decoder implementations such as \cite{higgott2023sparse,yuewu2022} are optimized primarily for patterns of Blossom/cluster creation encountered during decoding. The complementary gap method requires invoking a MWPM decoder on syndromes that differ appreciably from average case syndromes, leading to reduced effectiveness of optimizations \cite{newman2024private}.

Specifically, we construct two efficient approximate soft-output decoders for the surface code derived from the MWPM decoder and the UFD.
We prove analytical results about the prediction performance of the soft output and demonstrate numerically that the soft output accurately approximates the log-likelihood of successful decoding.
We then show that our soft-output decoder significantly outperforms the naive decoding of the hierarchical codes. 

We also analyze another application of soft-output decoding --- reducing the overhead cost of fault-tolerantly sampling the output distribution of a quantum circuit to specified accuracy. In this application, one runs a target circuit many times, and obtains a more reliable sample by discarding those runs in which the soft information signals that the probability of a logical error is intolerably large. 

In work by Gidney, Newman, Brooks, and Jones \cite{gidney2023yoked}, concatenating the surface code with the quantum parity code (stabilizer generators \(Z^{\otimes n}\) and \(X^{\otimes n}\)), known as the ``yoked surface code,'' has also been shown to result in physical qubit savings in the non-asymptotic regime when the inner surface code is decoded with a soft-output decoder similar to that of  \cite{hutter2014efficient}.
A surface code decoder that aborts when the estimated probability of a logical error is unacceptably high has been described very recently by Smith, Brown, and Bartlett~\cite{smith2024mitigating}.
Bomb{\'\i}n, Pant, Roberts, and Seetharam~\cite{bombin2024fault} have also used soft information to reduce the overhead of magic state distillation by abandoning those attempts at distillation where the probability of a logical error is high.

\begin{figure*}
    \centering
     \includegraphics[width=0.9\textwidth]{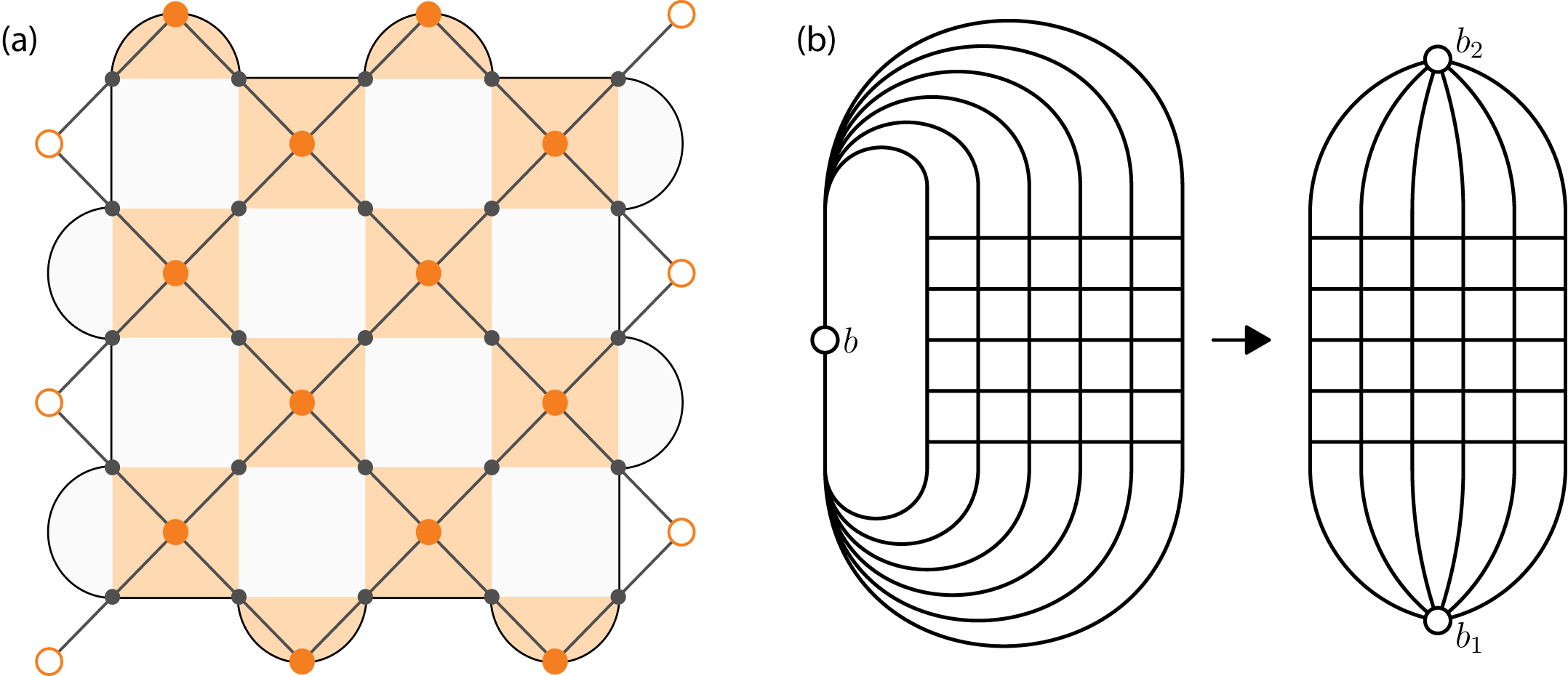} 
    \caption{(a) Creation of the decoding graph from a check matrix for the rotated surface code. Dark gray circles correspond to qubits while orange faces indicate the support of \(Z\) stabilizer generators.
    For every stabilizer generator, there is an orange vertex corresponding to an element in \(V_S\), and for every qubit, there is a corresponding edge. White vertices are identified with the boundary vertex \(b\). (b)
      Transformation of the decoding graph (of planar surface code) \(G_D=(V_S \sqcup \{b\}, E_D, \omega)\) to obtain the modified decoding graph \(G_D' = (V_S \sqcup B, E_D', \omega')\) with \(B=\{b_1,b_2\}\).
    }
    \label{fig:decoding_graph}
\end{figure*}

The paper is structured as follows. In \cref{sec:Background}, we provide necessary background and definitions for the proof and numerics. These definitions lay the groundwork for us to introduce our algorithm for extracting a soft output from MWPM and UFD. In \cref{sec:softinfo}, we present the modified decoding algorithms with soft output, prove some of their properties, and show numerically that the soft output quantity accurately captures the probability of a logical failure.
In \cref{sec:hc-decoding}, we apply the soft-output decoder to decode hierarchical codes, and in \cref{sec:sampling} we apply the soft-output decoding for improved sampling performance.

\section{Background}\label{sec:Background}
For a vector \(a \in \{0,1\}^n\) and Pauli operator \(U\), we write \(U^a\) to denote \(\prod_{i\in [n]}U_i^{a[i]}\).

A CSS code can be defined by its check matrices $H_X$ and $H_Z$.
The stabilizer generators are the Pauli operators $X^{\vec{a}}$ for each row $\vec{a}$ in $H_X$ and $Z^{\vec{b}}$ for each row $\vec{b}$ in $H_Z$. The CSS code is the simultaneous $+1$ eigenspace of these stabilizer generators.

With the exception of the application to the hierarchical code, we focus primarily on the surface code.
For a more comprehensive overview of surface codes, see \cite{dennis2002topological, fowler2012surface}

\subsection{Error and Syndrome}
In the depolarizing channel, errors are modeled as random bit flip ($X$), phase flip ($Z$) operators, or both ($Y = iZX$).
In this section, we will consider a stabilizer code on $n$ qubits.

\textbf{Syndrome}: A Pauli error $E \in \{I, X, Y, Z\}^n$ can be decomposed as $E \propto X^{\mathbf{e_x}} Z^{\mathbf{e_z}}$ where $\mathbf{e_x}, \mathbf{e_z} \in \{0,1\}^n$. Let $H_X$ and $H_Z$ be check matrices of a CSS code. Then, the syndrome is $
\begin{pmatrix}
    \sigma_x \\ 
    \sigma_z
\end{pmatrix}=
\begin{pmatrix}
    H_X \mathbf{e_z} \\
    H_Z \mathbf{e_x}
\end{pmatrix}$ over \(\F\). 

\textbf{Stochastic bit flip model}: In CSS codes, $X$ and $Z$ errors can be corrected separately, so we only consider the case of $X$ (bit flip) errors. For $n$ qubits, under stochastic bit flip noise with noise rate \(p \in [0,1]\), the applied operator is \(X^{\mathbf{e}}\) where for each \(i \in [n]\), \(\mathbf{e}[i] = 1\) with probability \(p\) and \(0\) with probability \(1-p\) i.i.d..

\subsection{Graph Definitions for QEC}
\subsubsection{Decoding Graph}

For a surface code with \(Z\) check matrix \(H_Z \in \F^{r \times n}\) and bit flip noise, we define a weighted graph, known as a decoding graph \(G_D=(V_D:= V_S \sqcup \{b\}, E_D, \omega)\) by associating a \emph{syndrome vertex} \(v \in V_S\) to every coordinate \(j\in [r]\) of the syndrome and an edge to every coordinate of the error \(i \in [n]\). The decoding graph of a rotated surface code is shown in \cref{fig:decoding_graph}a.
Each edge is incident to all vertices at which the corresponding error is detected.
I.e. \(i \in [n] \simeq E_D\) is incident to \(\supp H_Z[\cdot, i] \subseteq [r]\simeq V_S\).
Due to the topological nature of the surface code, the support is at most two.
We refer to errors detected at only one place as errors on the \emph{boundary} and add a special \emph{boundary vertex} \(b\) to the graph, so that the corresponding edge has two endpoints.

Later we will need the notion of \emph{inequivalent boundaries}.
We construct the modified decoding graph \(G_D'=(V_S \sqcup B, E_D', \omega')\) from \(G_D\) by replacing \(b\) with a set of vertices \(B\) such that a set of \(X\) errors corresponding to a path is a logical operator if and only if the endpoints are two distinct vertices in \(B\) with the edge weights induced by the weights in \(G_D\). This transformation from $G_D$ to $G_D'$ is shown in \cref{fig:decoding_graph}b. 
\(G_D\) is recovered from \(G_D'\) by identifying the vertex set \(B\) with a single vertex.
In other words, all cycles in \(G_D'\) correspond to a sum of \(X\) stabilizer generators of the code.
We say that two elements \(b,b' \in B\) are inequivalent boundaries if \(b\ne b'\).
Later, we will take the decoding graph to be a weighted graph where the weight of each edge corresponds to the log-likelihood marginal probability of the corresponding error occurring.

\subsubsection{Clusters}
A useful notion that we will use to extract a soft output signal is that of a \emph{cluster}.
For convenience, we define the notion of a cluster separately from any fixed error or syndrome pattern.
To a weighted graph \(G=(V,E,\omega)\) with all positive edge weights, we can associate a metric space \((X_G, d_G)\) where each edge is associated with an interval of length given by the edge weight (see \cref{sec:appendix_graph}) and vertices can be considered as points of \(X_G\).
The metric \(d_G \colon X_G \times X_G \to \mathbb{R}_+\) is given by the shortest path between two points.
We denote the measure on \(X_G\) by \(|\cdot|_\omega\).

\begin{defn}[Cluster]
    Given an assignment of radii \(\{r_v \in [0, \infty) \}_{v\in V}\) to the vertices of the graph, the corresponding cluster set is the union of balls around each vertex.
    In particular, 
    \begin{align}
        \mathcal{C}(\{r_v\}_{v\in V}) \equiv \bigcup_{v \in V} B_{r_v}(v) \subseteq X_G,
    \end{align}
    where \(B_{r_v}(v) \equiv \{x \in X_G : d(x,v) \le r_v\}\) is the ball of radius \(r_v\) centered on \(v\) using the canonical identification of \(V\) with points of \(X_G\).

    We refer to a connected component of a cluster set as a cluster.
\end{defn}
When clear from context, we suppress dependence on the radii set.

\subsubsection{Boundary elements}
Let \(G_D = (V_D, E_D, \omega)\) be a weighted decoding graph and let \(V_\sigma \subseteq V\) be the set of vertices for which the syndrome (of some arbitrary but fixed error) is nontrivial. To aid our proof with MWPM, we next define the boundary of a pair of vertices $e = (u,v)$ and a set of vertices $S$.

\begin{defn}[Boundary $\delta(e), \delta(S)$ and set $ \O$] Consider a fixed syndrome $\sigma$ and corresponding decoding graph $G_D=(V_D, E_D, \omega)$. 
Let $V_\sigma$ denote the set of vertices for which the error syndrome is nontrivial, and let $e$ be a pair $(u,v)$ of elements of $V_\sigma$. We will call a subset of $V_\sigma$ ``odd'' if it contains an odd number of elements. Let $\O$ denote the set of all the subsets of $\VNTS$ with odd cardinality. 

Let $S \in \O$ and let $E_{P, \sigma} = \{(u, v) \mid u \neq v; u,v \in V_{\sigma} \}$. We define $\delta(S)$ to be the set containing all pairs of elements of $V_\sigma$ such that $S$ contains exactly one element of the pair:
\begin{multline}
    \delta(S) := \{(u,  v) \in E_{P, \sigma} \mid(u \in S)\land (v \not\in S)\; \mathrm{or}\;  \\ 
     (u\not\in S)\land (v \in S) \}.
\end{multline}
For a vertex pair \(e \in E_{P, \sigma}\), we define $\delta(e)$ to be the set containing all odd subsets $T \in \O$ such that $T$ contains exactly one of $u$ or $v$, i.e. \(e\) is incident to \(T\): 
\begin{align}
    \delta(e) := \{T \in \O \mid e \in \delta(T)\}.
\end{align}

    \label{def:deltae}
\end{defn}
\noindent We note that $\delta(e),$ $\delta(S),$ and $\mathcal{O}$ are defined similarly in \cite{yuewu2022} and \cite{wu2023fusion}, except that we define them for the decoding graph.

The following definitions correspond roughly to matchings and edge weights in what is commonly known as the syndrome graph defined on the vertex set \(V_\sigma\) with edges corresponding to minimum weight paths.
While the minimum weight perfect matching decoder is most commonly defined on the syndrome graph, we find it most convenient for our purposes to work exclusively with the decoding graph. 

We begin by defining a valid edge set which intuitively corresponds to a valid correction.
However, we note that this set contains more information than a matching of the vertices of \(V_\sigma\) since there are multiple, topologically inequivalent ways to pair up two vertices of \(V_\sigma\).
\begin{defn}[Valid edge set]
\label{defn:validset}
    A valid edge set $E \subseteq E_D$ is one such that the number of edges in $E$ incident to a nontrivial syndrome vertex ($v \in \VNTS$) is odd, and the number of edges in $E$ incident to a trivial syndrome vertex ($v \notin \VNTS$) is even for all trivial syndrome vertices $v$. 
    We refer to an edge set as \emph{loop-free} if the induced subgraph is the disjoint union of path graphs.
\end{defn}

Here, we provide a notion of ``endpoints" of an edge set.
This should be thought of as the boundary mod-2 of a set of paths and corresponds with the usual boundary operator in the corresponding simplicial complex over \(\mathbb{Z}_2\) of the graph.
\begin{defn}[Edge set endpoints]
\label{defn:edge-set-endpoints}
    Given a valid loop-free edge set \(E \subseteq E_D\), it possesses a decomposition into a disjoint union of \(m\) sets \(E = E_1 \sqcup E_2 \dots E_m\) such that each \(E_i\) induces a path graph.
    Furthermore, each \(E_i\) has no vertices in common with the graph induced by the edge set \(E \setminus E_i\) i.e. the path graphs are pairwise disjoint.
    Let \(\tilde{\partial}\) be the map from an edge set to the pair of degree-1 vertices in the induced subgraph.
    Then, the endpoints of \(E\), denoted \(\partial E\), is defined to be the set \(\{\tilde{\partial}_i E_i\}_{i\in[m]}\).
    It is a set of pairs of vertices corresponding to the endpoint pairs of each path.
\end{defn}

\begin{defn}[Vertex pair weight]\label{defn:path-weight}
     For $e = (u,v) \in E_{P, \sigma}$, denote the minimal weight path between them by $P(u,v) \subseteq E_D$. The weight of $e$ is \begin{align}w_e = \sum_{q \in P(u,v)} |q|_\omega\end{align}
     where $|q|_\omega$ is the weight of the edge $q$ in the weighted decoding graph $G_D(V_D, E_D, \omega)$. 
\end{defn}
The minimal weight path is simply the shortest path with distance induced by the edge weights.
In the case of a square lattice with uniform edge weights, the distance of the shortest path coincides with the Manhattan distance.

\subsection{Decoder for the surface code}
A decoder is a map from the syndrome $\sigma$ to an error pattern $\mathbf{e}$ with a matching syndrome. We will consider the minimum weight perfect matching (MWPM) decoder \cite{dennis2002topological} and the union-find decoder (UFD) \cite{delfosse2021almost}.

\subsubsection{Minimum weight perfect matching}\label{sec:mwpm}
For noise channels with independent single qubit \(X\) and \(Z\) noise, MWPM solves for the most likely error \begin{equation}
    \mathbf{e} = 
    \argmax_{\mathbf{e}} 
    \Pr(\mathbf{e}|\sigma).
    \label{eq:mwpm}
\end{equation}
It does so by finding a minimum weight, valid set of paths $M$ in the weighted decoding graph $G_D(V_D, E_D, \omega)$, where a valid $M$ is defined in \cref{defn:validset}. When \(\omega\) is such that for each edge \(e\in E_D \), the corresponding error occurs with probability \(p_e\) and \(\omega(e) = \log\frac{1-p_e}{p_e}\), solving the MWPM problem corresponds to maximizing $\Pr(\mathbf{e}| \sigma)$ in \cref{eq:mwpm} \cite{edmonds1965maximum, kolmogorov2009blossom}.
In other words, MWPM returns the edge set \(E \subseteq E_D\) such that the endpoints match the syndrome and \(|E|_\omega\) is minimal \cite{dennis2002topological}.

\begin{fact}\label{fact:mwpm-edge-weights}
    Using the edge weights above, for any \(E\subseteq E_D\), we have that \(|E|_\omega = \mathrm{const.}-\log \Pr(E)\).
\end{fact}

The Blossom algorithm solves the MWPM problem by using the linear program (LP) formulation. 
For the decoding graph $G_D=(V_D, E_D, \omega)$, the MWPM problem is equivalent to an integer linear program (ILP), which can be relaxed to the following LP \cite{schrijver2003combinatorial}:
\begin{alignat*}{2}
  & \text{minimize: } & & \sum_{e \in E_{P, \sigma}} w_{e}x_{e} \\
   & \text{subject to: }& \quad & \begin{aligned}[t]
        \sum_{S \in \delta\left(\{v\}\right)} x_{e} & = 1, & \forall v & \in \VNTS \\
        \sum_{e \in \delta(S)} x_{e} & \geq 1, & \forall S & \in \O \text{ st. } |S| \geq 3\\
      x_{e} & \geq 0 & \forall e &\in E_{P, \sigma}
    \end{aligned}
\end{alignat*}
There is a primal variable $x_e$ for each pair of elements of $V_\sigma$, where $x_e = 1 \Leftrightarrow e \in M$ and $x_e = 0 \Leftrightarrow e \notin M$. 
It is known that including the redundant second constraint causes the solution of the integer program to be integral, and hence an optimal solution to this LP yields a minimum weight perfect matching \cite{schrijver2003combinatorial}.

We will also use the dual program of the MWPM LP. For each $S \in \O$ corresponding to a constraint of the primal LP, we define a dual variable $y_S$.
\begin{alignat*}{2}
  & \text{maximize: } & & \sum_{S \in \O} y_S \\
   & \text{subject to: }& \quad & \begin{aligned}[t]
        \slack(e) & \geq 0, & \forall e & \in E_{P, \sigma} \\
      y_S & \geq 0 & \forall S & \in \O
    \end{aligned}
\end{alignat*}
where $\slack(e) = w_e - \sum\limits_{S \in \delta(e)} y_S$. $e$ is tight if $\slack(e) = 0$.

The dual program provides a notion of clusters which we define in the following way \cite{yuewu2022,higgott2023sparse}.
\begin{defn}[MWPM Radius of a vertex]
    Given the decoding graph $G_D=(V_D, E_D,\omega)$ and feasible solution $y_S$ to the dual LP in the MWPM decoding problem, we define the radius $r_v$ of a vertex $v \in V_D$ as
    \begin{align}r_v = \sum_{S \in \O: v \in S} y_S.\end{align}
    \label{defn:mwpm-radii}
\end{defn}
\noindent The growth of clusters defined by these radii during the Blossom algorithm behaves similarly to the growth of clusters of Union Find \cite{yuewu2022} in that the final correction is supported within the clusters. This fact, which we will prove later, motivates \cref{defn:mwpm-radii}.

\subsubsection{Union Find Decoder} %
The Union-Find Decoder (UFD) \cite{delfosse2021almost} is a decoder for the surface code with almost linear time complexity. Despite better time complexity, the threshold of UFD is only slightly lower than that of MWPM.
We provide a brief review of necessary components of the UF decoder.
For a full introduction, readers should consult \cite{delfosse2021almost}.

UFD operates on the decoding graph $G_D=(V_D, E_D, \omega)$, as opposed to the syndrome graph.
On this graph $G_D$, UFD first generates clusters such that a valid correction operator is contained within the clusters.
These clusters are grown by expanding around the nontrivial syndrome vertices until either an even number of these syndrome vertices are in the cluster, or the cluster touches the boundary.
The clusters have the interpretation of erasures in the sense that the first stage of the decoder attempts the easier task of identifying a subset of the qubits on which the true error is supported.

The second stage finds a valid correction operator supported within the clusters.
This task is identical to correcting erasures and can be done efficiently by the peeling decoder (depth-first search traversal).
If the clusters are topologically trivial, then any correction contained within the support will suffice.

We now motivate the precise definition of the soft output quantity for UFD:
Since the clusters are treated as erasures in the second stage, it is reasonable to consider the second-stage clusters as erasure errors along with undetected Pauli errors. Then, the probability of a logical error should be thought of as the probability that an undetected Pauli error connects the two boundaries.
Without conditioning on the syndrome or accounting for degeneracies, this corresponds to the minimum weight path between the boundaries in the decoding graph after setting edge weight corresponding to erased edges to zero.
The precise support of the logical operator within the erased clusters is not important as the clusters are individually correctable, and any correction is equivalent.

We now define UFD on the metric space $X_G$ induced by $G_D$ \cite{huang2020fault,pattison2021improved}. 
An odd cluster is a connected component of the cluster set $\mathcal{C}(\{r_v\}_{v \in V})$ such that there are an odd number of nontrivial syndrome vertices within the cluster. 
For convenience, we assume that all edges have the weight \(w\).

\RestyleAlgo{ruled}
\begin{algorithm}
\caption{Union-Find decoder}\label{algo:ufd}
\KwIn{Decoding graph $G_D=(V_D, E_D, \omega)$ (with uniform edge weights \(w\) assumed for simplicity), an error $\mathbf{e}$, and it's corresponding syndrome $\sigma \subset V_D$.}
\KwOut{A correction $F\subset E_D$ for error $\mathbf{e}$.}
\BlankLine
\ForAll{nontrivial syndrome vertex $i$}{initialize radius $r_i \gets 0$}
\While{\(\exists\) odd clusters in the cluster set \(\mathcal{C}\left(\{r_i\}_i\right)\)}{
    \ForAll{nontrivial syndrome vertices $i$ contained in the smallest odd cluster}{
    \label{algo:ufd:increment}
    $r_i \gets r_i + \frac{w}{2}$\;
    }
}
\end{algorithm}
\noindent Note that \cref{algo:ufd:increment} may cause odd clusters to become even.
All odd clusters should have been visited (to increment the radius) before returning to a previously visited odd cluster.

\begin{defn}[UFD Radius of a vertex]\label{defn:ufd-radii}
Given the decoding graph $G_D=(V_D, E_D, \omega)$, the radii of a non-trivial syndrome vertex $v \in V_D$ from running UFD are the final $r_i$ from Algorithm \ref{algo:ufd}.
\end{defn}

\subsection{Hierarchical Code}
A concatenated code combines two codes, an inner code $C_1$ and an outer code $C_2$. If the inner code encodes only 1 logical qubit, this concatenated code, is created by encoding each qubit of the code $C_2$ into a copy of $C_1$.

The hierarchical code is a concatenated code where the inner layer is a topological code, such as a surface code or color code, and the outer layer is a constant-rate quantum LDPC code \cite{pattison2023hierarchical}. This specific choice of codes permits a threshold when restricted to geometrically local gates in two dimensions.
Here, we will always take the outer code to be the particular Quasi-cyclic Lifted Product Code (QCLP) \cite{pantaleev2021quantum} described in \cref{sec:qclp-codes} which encodes \(140\) logical qubits into \(1054\) physical qubits. From numerics in \cite{pantaleev2021quantum}, its distance is believed to be about $d=20$.

A naive decoder for a concatenated code first decodes the inner code and then the outer code. However, Poulin showed \cite{poulin2006optimal} significant improvements to the decoding performance if the inner and outer codes are decoded jointly.
Additionally, it is known that iterative message-passing decoders perform poorly on quantum LDPC codes due to the short cycles in the Tanner graph arising from degeneracy of the quantum code \cite{poulin2008iterative}.
Soft information arising from the decoding of the inner code provides a natural means to break the degeneracy and improve the performance of belief propagation on the outer code.

\begin{figure*}[ht]
    \centering
    \includegraphics[width=\textwidth]{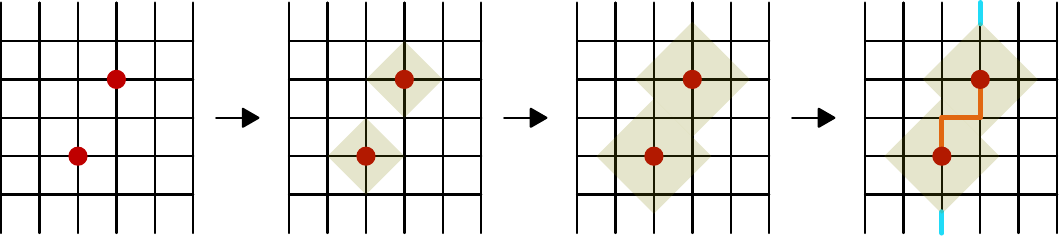}
    \caption{
    Extracting soft output. Here, we depict the decoding graph of a $d=5$ planar surface code with an example syndrome in red.
    To extract the soft output, we take the fully grown clusters $\mathcal{C}$ from UFD or MWPM (in beige) and find the minimum path (blue) between inequivalent boundaries, considering edges within the cluster to be of weight 0. The soft output $\phi(\mathcal{C})$ is the length of the path in blue.
    This example considers the 2D decoding problem for perfect syndrome measurements for simplicity, but the same procedure applies to a 3D decoding graph corresponding to faulty measurements. 
    }
    \label{fig:phi_path}
\end{figure*}

\section{\label{sec:softinfo}Soft-output decoding}
In this section, we introduce an efficient method to convert MWPM and UFD to supply a soft output. 
In the setting of the classical repetition code, we can prove that this soft output is equal to the log-likelihood of a logical failure.
However for surface codes, we are only partially able to prove the relationship between the soft output and the log-likelihood of a logical fault, so we supplement our understanding with numerics.

\subsection{Extracting the soft output}
UFD converts the case of depolarizing noise to that of erasure by growing clusters that cover the error.
If the clusters cover the error and the clusters themselves are topologically trivial then any valid correction will result in a trivial residual error.
In order for an error configuration to result in a logical fault, an error chain must cover a path between two or more clusters such that the union of the cluster and the error chain cover a logical operator. 
Thus, the distance between clusters should contain some information about how likely the decoder is to have succeeded.
 To quantify this, we define a quantity $\phi$ based on a set of clusters \(\mathcal{C}\) as follows:

\begin{defn}[Soft output for surface codes]\label{defn:soft-info}
    Given a set of clusters \(\mathcal{C}\), there exists a decomposition into \(r\) connected components \(\mathcal{C} = \sqcup_{i\in [r]} \mathcal{C}_i\).
    Define a new metric space $(X_{G_D}', d_{G_D}')$ that is the quotient of $(X_{G_D}, d_{G_D})$ by each \(\mathcal{C}_i\) i.e. identify each \(\mathcal{C}_i\) with a point.
    Define $\phi(\mathcal{C})$ to be the length of the shortest path that covers a logical operator in $X_{G_D}'$.
\end{defn}

\RestyleAlgo{ruled}
\begin{algorithm}
\caption{Soft output in surface codes}\label{algo:softinfo}
\KwIn{The weighted decoding graph $G_D(V_D, E_D, \omega)$, 2 inequivalent boundary nodes $b_1, b_2$, and the radii assignments $\{r_v\}$ after running MWPM or UFD.}
\KwOut{Soft output $\phi(\mathcal{C})$.}
\BlankLine
Create new graph $G_D'(V_D, E_D, \omega')$ where any edges within $\mathcal{C}(\{r_v\})$ have edge weight 0.  \\
Run Dijkstra's algorithm on $G_D'$ with source vertex $b_1$ to find $\phi(\mathcal{C})$, the minimum path length from $b_1$ to $b_2$.
\end{algorithm}

As in Algorithm \ref{algo:softinfo}, for surface codes and toric codes, \(\phi\) can be efficiently computed using Dijkstra's algorithm, with runtime $O(|E_D|+|V_D| \log |V_D|)$.
When using the clusters generated by a particular decoding algorithm, we will notate \(\phi(\mathcal{C})\) as a function of the syndrome \(\sigma\) i.e.  \(\phi(\sigma)\).

Intuitively, for a set of clusters, \(\phi(\sigma)\) is roughly a lower bound on the log-likelihood of an error chain that is inequivalent to corrections within the clusters. 
Unfortunately, this statement is somewhat difficult to show due to an insufficient analytical handle on the structure of the correction generated by MWPM or UFD within the clusters. 

\subsection{Analytics}

In this section, we provide analytical evidence that \(\phi(\sigma)\) is a good approximation to the log-likelihood of a logical fault after decoding.

\subsubsection{UFD and Repetition Code}
\label{sec:ufd-repcode}
We begin by proving that our soft output quantity is the log-likelihood of a decoding failure when applied to UFD on the classical repetition code.

The classical repetition code of length \(n\) has check matrix given by \(n-1\) rows where the \(i\)-th row has a \(1\) only on the coordinates \(i\) and \(i+1\).
The corresponding decoding graph \(G_D = (V_D,E_D,\omega)\) is the cycle graph on \(n\) vertices where the boundary vertex can be made ``real" by adding an additional redundant row to the check matrix supported on the first and last bits.
We assume this has been done.
For i.i.d. bit flip noise with bit flip rate \(p\), we assign the edge weight \(\log \frac{1-p}{p}\) to all edges corresponding to the log-likelihood of a bit flip.

For this code, both MWPM and UFD correct all errors of weight \(t \le \lfloor\frac{n-1}{2}\rfloor\).
For \(n\) odd, this means that both decoders return the same correction.
However, we find the result in the case of UFD to be easier to prove. For MWPM, establishing a connection between the 1D structure of the decoding graph and the resulting structure of the dual solution is somewhat challenging for reasons similar to why we find it difficult to establish an approximate upper bound in \cref{thm:mwpm_surfcode} of \cref{sec:mwmp-surface-code}.

\begin{theorem}
    Consider a repetition code on an odd number $n$ of data bits, error $E$ distributed according to a i.i.d. stochastic bit flip model with error probability $p$. Let \(G_D= (V_D,E_D,\omega)\) be the decoding graph with uniform edge weights \(w = \log \frac{1-p}{p}\), $\sigma$ be the syndrome of $E$, $F$ be the set of edges in the correction produced by UFD.
    Lastly, let $\phi(\sigma)$ be the soft output quantity computed from the output of UFD according to \cref{defn:soft-info}.
    Then, 
    \begin{align}
        \log \frac{\Pr(E = F | \sigma)}{\Pr(E \ne F | \sigma)} = \phi(\sigma)
    \end{align}
    \label{thm:ufd_repcode}
\end{theorem}
\begin{proof}
  Since the distance of the repetition code is equal to \(n\) and UFD corrects all errors of weight less than \(d/2\) \cite{delfosse2021almost}, UFD will always return the minimum weight correction \(F\) to the syndrome \(\sigma\).
  The other valid correction to \(\sigma\) is \(F^c\) which has weight \(n-|F|\).
  Since \(n\) is odd, \(E\ne F\) if and only if \(|E| > \frac{n}{2}\).
  Thus,
  \begin{align}
    \frac{\Pr(F = E | \sigma)}{\Pr(F \ne E | \sigma)} &= \frac{p^{|F|} (1-p)^{n-|F|}}{p^{n-|F|} (1-p)^{|F|}} \\
    &= \left(\frac{1-p}{p}\right)^{n-2|F|}
    \label{thm:ufd_repcode:n_2f}
  \end{align}

  It remains to compute \(\phi\) for the syndrome \(\sigma\).
  For the repetition code, the only non-trivial logical operator covers all edges, so \(\phi\) is the weight outside of the clusters \(C\) i.e. \(\phi = |C^c|_\omega = w n - |C|_\omega\).
  We proceed by using an argument very similar to the correctness proof of UFD \cite[theorem 1]{delfosse2021almost}.
  During cluster growth, using the decoding graph of the repetition code, each connected cluster is odd if and only if every valid correction to the syndrome contains exactly one path of edges leaving the cluster: Otherwise, the cluster could not contain an odd number of non-trivial syndrome vertices.
  Thus, each time a cluster is grown, the intersection \(|C \cap F|\) is increased by \(\frac{1}{2}\) while \(|C|\) is increased by \(1\); \(\frac{1}{2}\) for each of the two frontiers.
  After \(2|F|\) growth steps, \(F\) is completely contained in the clusters, so the growth process must halt.
  We conclude that \(|C|_\omega = 2|F|_\omega = 2 w |F|\) and so \(\phi = (n-2 |F|) w \).

  Returning to \cref{thm:ufd_repcode:n_2f}, we find
  \begin{align}
      \log \frac{\Pr(F = E | \sigma)}{\Pr(F \ne E | \sigma)} &= \left( n-2|F|\right)\log\left(\frac{1-p}{p}\right) \\
      &= \phi
  \end{align}
\end{proof}

Crucially, the bound here relies on two facts: The minimum weight correction in the opposite equivalence class is known and the intersection with the clusters can be computed in terms of the minimum weight correction \(F\) due to the 1D nature of the decoding graph.
While one might hope for a partial result for UFD in the setting of surface codes, the characterization of cluster growth in UFD only holds for errors up to weight \(d/2\) which presents a further obstruction.
Perhaps an analysis based on the fact that the correction to far-separated errors is independent may help; however we leave this to further work.

\subsubsection{MWPM and Surface Code}
\label{sec:mwmp-surface-code}
We now turn our attention to the surface code where we will prove a lower bound on the relation between $\phi(\sigma)$ and log probability ratio of minimal-weight errors in the two different equivalence classes for MWPM decoding.
In that computing the most likely error approximates the most likely equivalence class \cite{wang2003confinement}, we believe that this approximates the log-likelihood of a decoding failure.
It is also worth noting that the main result of this section, \cref{thm:mwpm_surfcode}, holds as well for the repetition code of the previous section.

We begin by proving two lemmas which allow us to show that the intersection of any valid edge set with the clusters must be at least that of the minimum weight solution.
Intuitively, while we have not proved it, the optimal correction is completely contained within the clusters, so we would like to quantify how suboptimal a given valid edge set is by how much of it is outside of the clusters.

The heart of the argument is that any valid correction \(M\) must intersect with the clusters \(C\) at least as much as the minimum weight correction.
We begin by proving two lemmas establishing this fact via the dual variables.

\begin{lemma} For any valid loop-free set of edges $M \subset E_D$ on the decoding graph $G_D=(V_D, E_D,\omega)$,  and any optimal solution $y_S$ to the dual LP and its corresponding clusters $\mathcal{C}$,
    \begin{align}
      |M \cap \C|_\omega \geq \sum_{e \in \partial M} \sum_{S \in \delta(e)} y_S
    \end{align}
    \label{claim:cluster_is_ys}
\end{lemma}
\begin{proof}
    For an arbitrary path $p \subseteq E_D$ connecting $u,v \in V_D$, define \(e = \partial p = (u,v)\).
    By the definition of clusters as the union of balls,
\begin{align}
|p \cap C|_\omega &\geq \min(w_e, r_u + r_v) \\
&\geq \min\left(w_e, r_u + r_v - 2 \sum_{S \in \O: u, v \in S} y_S\right) \\
&= \min\left(w_e, \sum_{\substack{S \in \O: \\ u \in S}} y_S + \sum_{\substack{S \in \O: \\ v \in S}} y_S - 2 \sum_{\substack{S \in \O: \\ u, v \in S}} y_S\right) \\
&= \min\left(w_e, \sum_{S \in \delta(e)} y_S\right)
\end{align}
Where we substitute the definition of the radius and combine the sums using \cref{def:deltae} that \(\delta(e) = \delta((u,v)) = \left\{S \in \O : (u\in S)\land (v \not\in S)\; \mathrm{or}\; (u\not\in S)\land (v \in S)\right\}\).

The dual constraints (\cref{sec:mwpm}) require that $\slack(e) \geq 0$, so $\sum_{S \in \delta(e)}y_S \leq w_e$.
We can further simplify the min to 
\begin{align}
    |p \cap C|_\omega \geq \sum_{S \in \delta(e)}y_S
\end{align}

Returning to the bound, let \(
\{M_i\}_{i \in [m]}\) be a path decomposition of \(M\) into \(m\) disjoint paths.
We apply the previous lower bound for each path in the path decomposition of \(M\).
\begin{align}
    | M \cap C|_\omega &= \sum_{i \in [m]} |M_i \cap C|_\omega \\
    &\geq \sum_{e \in \partial M} \sum_{S \in \delta(e)}y_S
\end{align}

\end{proof}

\begin{lemma} For any valid loop-free set of edges $M$ in the decoding graph $G_D=(V_D, E_D,\omega)$, optimal solution $y_S$ to the dual, and solution to the MWPM problem $F \subseteq E_D$,
\begin{align}
    \sum_{e \in \partial M} \sum_{S \in \delta(e)} y_S \geq |F|_\omega
\end{align}
\label{claim:ysG_equal_ysF}
\end{lemma}
\begin{proof}
First, note that for any valid loop-free set of edges $M$, 
\begin{align}
    \sum_{e \in \partial M} \sum_{S \in \delta(e)} y_S &= \sum_{e \in \partial M} \sum_{S \in \O: e \in \delta(S)} y_S\\
    &= \sum_{S \in \O} y_S |\partial M \cap \delta(S)|.
    \label{eq:schrivjermagic}
\end{align}
For a given \(S \in \O\), the number of times \(y_S\) appears in the sum is the number of intersections \(\partial M \cap \delta(S)\) i.e. the number of times a vertex pair \(e \in \partial M\) is incident to \(S\) \cite[Pg. 454]{schrijver2003combinatorial}.

Because $S$ is of odd cardinality, $|\partial M \cap \delta(S)| \geq 1$,
\begin{align}
\sum_{e \in \partial M} \sum_{S \in \delta(e)} y_S &= \sum_{S \in \O} y_S |\partial M \cap \delta(S)| \\
&\geq \sum_{S \in \O} y_S \\
&= |F|_\omega
\end{align}
where the last simplification is because $|F|_\omega$, the solution to the LP, achieves the maximum of the objective in the dual LP.

\end{proof}

We are now ready to prove a relation between \(\phi\) and the log-likelihood ratio of the most likely errors in the two equivalence classes.
Having shown that the edge set corresponding to the opposite equivalence class must have at least weight inside of the clusters as the optimal edge set, we employ the definition of \(\phi\) to lower bound the weight \emph{outside} the clusters.
\begin{theorem}
    Consider a surface code on $n$ data qubits, error $E$ distributed according to a stochastic bit flip model with error probability $p$. Let $\sigma$ be the syndrome of $E$, $F$ be the set of edges in the correction produced by MWPM, $M$ be the minimum weight, valid set of edges in the opposite logical class i.e. the symmetric difference of \(M\) and \(F\) is a non-trivial logical operator.
    Lastly, let $\phi(\sigma)$ be the soft output quantity computed from the output of MWPM according to \cref{defn:soft-info}.
    Then, 
    \begin{align}\log \frac{\Pr(E = F | \sigma)}{\Pr(E = M | \sigma)} \geq \phi(\sigma)\end{align}
    \label{thm:mwpm_surfcode}
\end{theorem}

\begin{proof}
   Since \(M\) and \(F\) are both minimal in their respective equivalence classes, they are valid loop-free edge sets.
   $M$ can be split into the region inside the clusters and outside
   \begin{align}
       M = (M \cap \C) \sqcup (\C^c \cap M).
   \end{align}
   By the definition of $\phi$ and $M$ we also have that
   \begin{align}
     |\C^c \cap M|_\omega \geq \phi(\sigma).
   \end{align}
   
   We now lower bound \(|M|_\omega - |F|_\omega\),
   \begin{align}
    |M|_\omega - |F|_\omega &= |M \cap \C|_\omega - |F|_\omega + |\C^c \cap M|_\omega \\&\geq |M \cap \C|_\omega - |F|_\omega  + \phi(\sigma) \\
    &\ge \phi(\sigma)
   \end{align}
   Where we have used lemma \ref{claim:cluster_is_ys} and lemma \ref{claim:ysG_equal_ysF} to lower bound \(|M \cap C|_\omega \geq |F|_\omega\).
  Using this bound, \cref{fact:mwpm-edge-weights}, and Bayes' rule,
  \begin{align}
    \log \frac{\Pr(E = F|\sigma)}{\Pr(E = M|\sigma)} &= 
    \log\frac{\Pr(E = F)}{\Pr(E = M)} \\
    &= |M|_\omega - |F|_\omega\\
    &\ge \phi(\sigma)
  \end{align}
\end{proof}

Given the results of the numerics, a natural question is whether \(\phi(\sigma)\) also approximately upper bounds the log-likelihood ratio i.e., does there exist a constant \(C \ge 1\) such that \(\log \frac{\Pr(E = F|\sigma)}{\Pr(E = M|\sigma)} \le C\phi(\sigma)\)?
Unfortunately, we were not able to establish any such bound in the setting of surface codes due to an insufficient handle on the correction in the opposite equivalence class.
In particular, \(M\) may not even correspond to an output of MWPM, so many of the properties of the dual program are lost. 
Furthermore, our intuition regarding \(\phi(\sigma)\) is that the minimal weight path between clusters is the amount of ``excess'' weight \(M\) must have.
However, it is difficult to establish many properties of such a valid edge set: Which vertices should be used to match between clusters? Once a vertex pair has been removed from a cluster, can the remaining vertices be paired up without significantly increasing the weight of the solution?

\begin{figure*}[]
    \centering
    \includegraphics{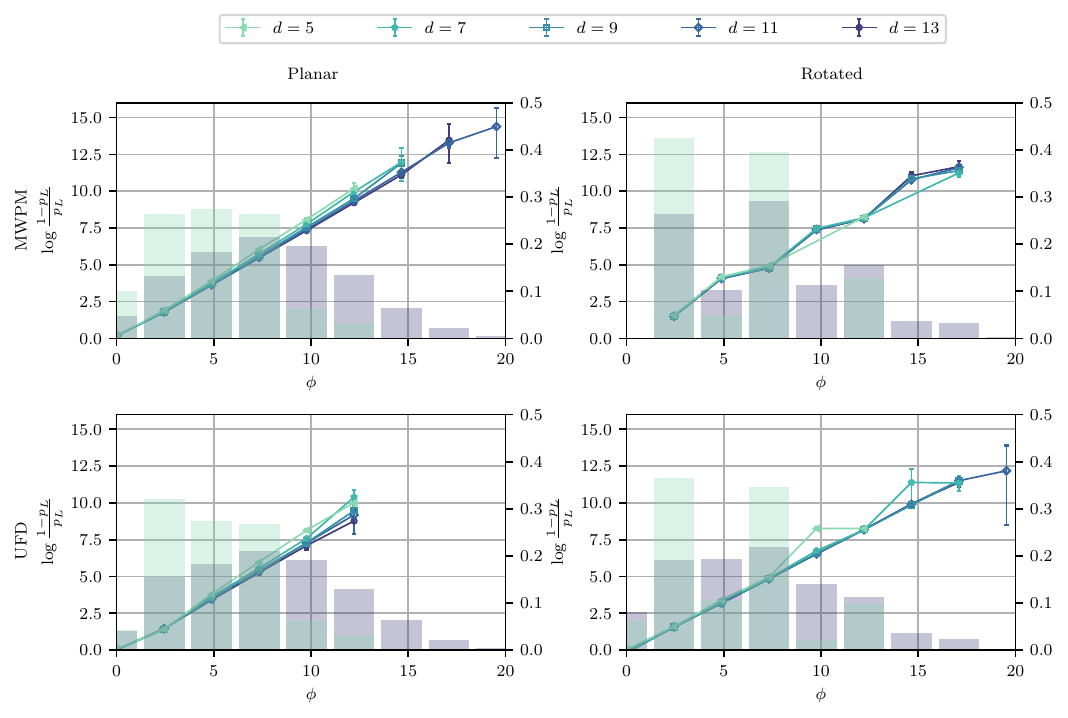}
    \caption{Correlation between soft output signal $\phi$ (\cref{defn:soft-info}) and the log-likelihood ratio of $p_L= \Pr(\text{decoder returned a logical failure} | \sigma)$ for individual physical qubit error rate  $p = 0.08$ and surface code distance $d$, assuming syndrome measurement is perfect.
    A line is drawn as a guide to the eye.
    Note that a log-likelihood value of \(12\) corresponds roughly to a failure rate of \num{6e-6}.
    We show (a) planar surface code with MWPM, (b) rotated surface code with MWPM, (c) planar with UFD, and (d) rotated with UFD. We plot the marginal sample distribution of $\phi$ as green and purple bar charts (mass on right $y$ axis) with $d=5$ and $d=13$ respectively. With the exception of the planar surface code run with UFD, all combinations of parameters plotted have between $10^7$ and $10^9$ samples. For UFD on the planar surface code, there are fewer ($10^5 - 10^7$) samples due to runtime constraints. 
    Error bars indicate 95\% confidence intervals.
    Markers for \(\phi\) values where no failure was observed (infinite sample log-likelihood of success) are omitted.
    For larger $d$, the tail of the \(\phi\) distribution extends beyond \(22\) up to a maximum of \(30\). For $d=11$ and $d=13$, approximately $1$ event in $10^4-10^6$ have $\phi$ larger than 22, but no failures were observed for these large $\phi$ values.
    }
    \label{fig:phi_linear}
\end{figure*}
\subsection{Numerics}

\begin{figure}[]
\centering
\includegraphics{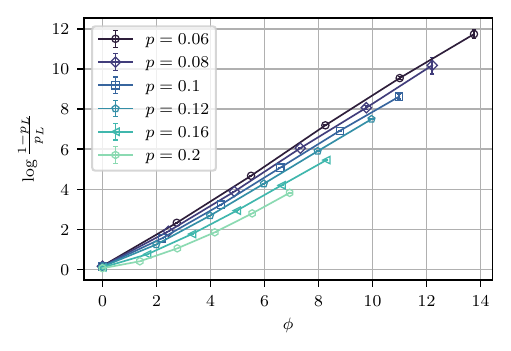}
\caption{Correlation of soft output $\phi$ with the probability $p_L$ of a logical error in the distance-5 planar surface code for various values of the physical error rate $p$, assuming perfect syndrome measurement.
We note that the linear relationship is less apparent for $p$ above the MWPM threshold ($p \sim 0.103$).
However, it is still present.
} 
\label{fig:varyp_dfixedd}
\end{figure}

We first simulated the surface code under bit flip noise and assuming perfect syndrome extraction. We find that, conditioned on the syndrome, $\phi$ is well correlated with the log-likelihood of a logical error.

In \cref{fig:phi_linear}, we show the linear relationship for combinations of MWPM/UFD and rotated/planar surface codes.
Notably, while the planar surface code has nearly perfect agreement within sampling error, the rotated surface code exhibits some small ``artifacts."
These could be due to the somewhat less uniform lattice and different entropic factors. 

In \cref{fig:varyp_dfixedd}, we demonstrate that the linear relationship holds for different values of the physical error rate by considering a \(d=5\) surface code decoded using MWPM.
We observe a linear relationship for all physical error rates $p$ below MWPM's threshold \(p = 10.3\%\) for perfect syndrome measurement \cite{dennis2002topological, wang2003confinement}.
We observe similar results in the presence of measurement errors.

Notably, in nearly all cases, the slope of \(\phi\) vs the log-likelihood of a logical error is almost constant with respect to \(d\) and \(p\), as the probability ranges over about 6 orders of magnitude.
This numerical evidence strongly suggests that 
\(\phi\) closely tracks the log-likelihood of a logical error.

To perform these simulations, we used a combination of custom code and open source packages, including the Fusion Blossom package \cite{wu2023fusion}, qsurface \cite{qsurface}, Stim \cite{gidney2021stim}, and the ldpc library \cite{Roffe_LDPC_Python_tools_2022}.

\section{Decoding the Hierarchical Code using soft information}
\label{sec:hc-decoding}

In this section, we present our numerical results from using soft information to decode the hierarchical code.
In designing a decoder for the hierarchical code, we can use either MWPM or UFD on the inner surface codes and calculate the soft output signal.
In the context of decoding the outer code, the soft output from the inner code is related to the notion of \emph{soft information}.
Thus, we refer to the soft output as soft information when it is being used in the decoding of concatenated codes.
Given the soft information, we must next develop a decoder for the outer code that uses this soft information.
Belief propagation (BP) is a promising proposed decoder in the literature for qLDPC codes \cite{poulin2008iterative}. 
It passes messages between the physical qubits and check bits, repeatedly updating the marginal ($\Pr(\text{bit failed})$) of the qubits until all checks are satisfied.

BP performs well on classical LDPC codes, but frequently fails to decode quantum LDPC codes due to degeneracy:
There are multiple errors with the same syndrome that differ only by a low-weight stabilizer generator.
BP cannot distinguish these and ends up stuck in a local minimum.

To mitigate this issue, various modifications to BP have been proposed (see for example \cite{grospellier2021combining, liu2019neural, poulin2008iterative, raveendran2019syndrome, kuo2020refined, du2022stabilizer, kuo2022exploiting}).
One solution is to randomly perturb the priors and thus break the degeneracy \cite{poulin2008iterative}.
In the setting of concatenated quantum codes \cite{poulin2006optimal}, it was shown by Poulin that decoding an outer code using a prior obtained from the inner code greatly improves the error correction performance.
When the outer code is decoded using BP, a soft-output decoding of the inner code naturally provides a non-uniform prior that reduces the effects of degeneracy.
We thus expect that
using soft information as a prior in BP can greatly improve performance.

\subsection{Simulation methods}
\label{sec:sim-methods}
We use sum-product belief propagation (BP) \cite{grospellier2021combining} to decode the outer code.
Instead of jointly simulating the inner and outer code, we first sample the joint distribution of \(\phi\) and the presence of an inner-code logical failure using UFD. 
To sample errors/soft information for the outer code, we sample from the empirical joint distribution of soft information signal and inner code logical error $\Phi$. 
The prior supplied to BP is the empirical failure rate conditioned on the sampled value of \(\phi\).
In this two-step approach, we use \(10^6\) samples to construct the empirical distribution $\Phi$.
The two-step approach has the drawback that it is unclear how to compute the sampling error, so we do not attempt to compute an error bar for these numerics.
In general, we expect failures of the outer code to arise from a combination of many common events instead of few rare events, so the error incurred by the two-step approach should be small.
We provide evidence that the joint distribution is adequately sampled in \cref{app:sampling}.

All simulations of the outer code use \(100\) rounds of syndrome extraction under bit flip noise and measurement noise sampled from $\Phi$.
These 100 syndrome extraction rounds are followed by one round of perfect syndrome extraction. In other words, with the exception of the final syndrome extraction round, bit flip noise with a soft signal is applied to the data and measurements  after each syndrome extraction round, where the bit flip probability and soft information signal are sampled from the empirical joint distribution \(\Phi\) i.i.d. 

Even though the final syndrome extraction is perfect, it is not guaranteed that the BP decoder returns the state to the code space. Therefore, to determine the logical error rate, we declare a failure if and only if a predetermined basis of logical operators fails to commute with the residual error.
We run BP on the full syndrome history after taking the difference of syndromes in consecutive rounds \cite{dennis2002topological} with a flooding schedule. A flooding schedule is one where all nodes update their messages simultaneously and send out new messages to their neighbors all at once, rather than in a sequential or staggered manner. 
We include brief background on BP in \cref{app:belief-propagation} and refer readers to \cite{richardson2008modern} or \cite{poulin2008iterative} for a full description.

\subsection{Pseudothreshold using soft information}
\begin{figure}[]
    \centering
    \includegraphics{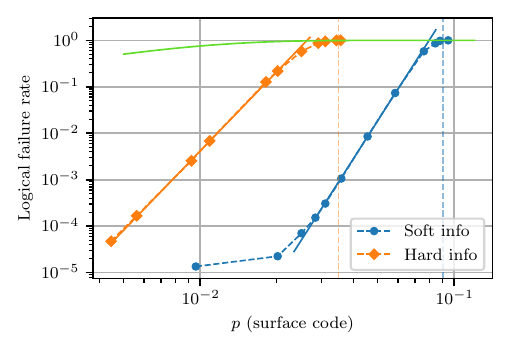}
    \caption{
    Logical failure rate for the hierarchical code, decoded with (in blue) and without (in orange) soft information, as a function of the logical error rate of the inner surface code. Shown in green is the probability that in a set of $k$ surface codes, at least one of them fails. Vertical dotted lines are the pseudothresholds for each decoding method.
    The inner code is a distance \(5\) rotated surface code, with $T=49$ syndrome extraction rounds.
    }
    \label{fig:pseudothreshold}
\end{figure}

To quantify how the soft information improves the effectiveness of decoding the outer code, 
we plot the performance of a hierarchical code in terms of the failure rate of a small inner surface code (\(d=5\)) instead of the physical error rate.
Using a small inner code makes it easier to generate a large enough sample to accurately estimate the joint distribution for $\phi$ and the inner-code logical error probability. 
We note, though, that a smaller inner code yields less informative soft information than a larger inner code, so we expect soft information to be less advantageous for a smaller inner code. In this sense, choosing $d=5$ underestimates the gain in performance that can be achieved by choosing a larger inner code.

In \cref{fig:pseudothreshold}, we show that including soft information improves the pseudothreshold of the outer QCLP code by about a factor of 3. To obtain the joint distribution of inner-code logical errors and \(\phi\), the surface code is simulated with faulty measurements and bit flip errors for 49 rounds of syndrome extraction. 
With soft information, we provide a prior as in \cref{sec:pseudothreshold_sim}, whereas with hard information, we only provide a uniform prior given by the marginal logical error rate.
For the outer code, we use the simulation method with noisy measurements as described in \cref{sec:sim-methods}. Here, both bit flip errors and measurement errors in the outer code are all sampled from the inner code's soft information joint distribution where inner-code syndrome measurement errors are included.

By adding soft information, the pseudothreshold of the outer code improves from roughly \(p_\mathrm{surface~code}=0.032\) to \(0.09\), where \(p_\mathrm{surface~code}\) refers to the logical failure rate of the inner surface code. We consider the logical error rate of the outer code as a function of the logical error rate of the inner code (rather than the physical error rate) in an effort to assess the benefit of using soft information without making direct reference to how the inner code is chosen. %
In addition, when soft information is included, we find that the probability of an outer-code logical error scales more favorably as a function of the inner-code error rate. Fitting to a power law $\propto p_\mathrm{surface}^\alpha$ in the below-pseudothreshold regime before the error floor, we find $\alpha = 8.5(1)$ and  $\alpha = 5.66(2)$ for decoding with and without soft information respectively. 

\subsection{Hierarchical code performance}
\label{sec:pseudothreshold_sim}

\begin{figure*}
    \centering    \includegraphics{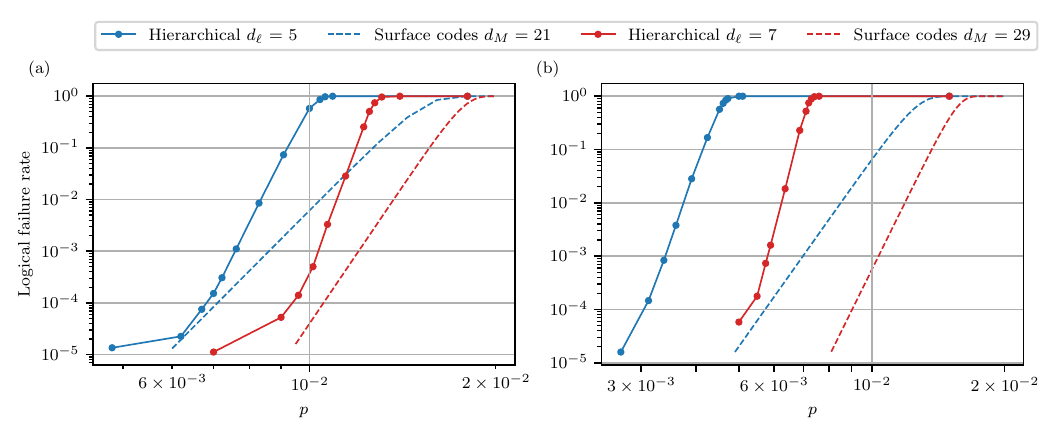}
\caption{Comparison of the logical failure rate for hierarchical code memory (solid) vs surface codes (dashed) storing the same number of logical qubits with equivalent number of physical qubits. 
    For each individual physical qubit error rate $p$, we simulated the hierarchical logical failure rate under phenomenological bit flip noise with faulty measurements. The hierarchical code (solid line) consists of an outer qLDPC code that is a quasi-cyclic lifted product code with parameters $\llbracket 1054, 140, \_ \rrbracket$. The inner code is a surface code with distance $d_\ell = 5$ or $7$. The surface codes for comparison (dashed lines) have distance $d_M=21$ or $29$. $r=10$ in (a) and $r=1$ in (b), where $1/r$ is the ratio of SWAP/idle error rate to the error rate of all other gates. We sample over $T = \lceil 3 \sqrt{n}d/r \rceil$ syndrome extraction rounds, corresponding to $T=49$ and $T=487$ for $r=10$ and $r=1$ of the $d=5$ (solid blue) surface codes respectively, and $T=69$ and $T=682$ for the $d=7$ (solid red) surface codes.
    The logical error rate floor for the hierarchical code encountered at low physical error rates is discussed at the end of \cref{sec:pseudothreshold_sim}.}
\label{fig:hierarchical_surfcode_comparison}
\end{figure*}

To estimate the error correction performance of the hierarchical code, we again take a two-step approach --- first we sample the joint error distribution of errors and soft output for the inner surface code, then we use that distribution to analyze the performance of the outer code.
However, we must account for the failure rate of the SWAP gates in the deep syndrome extraction circuit of the hierarchical code.
Since the objective was to simulate the hierarchical code under an analog of phenomenological noise with data qubit bit flips and measurement errors, we account for the error of the SWAP gates in this spirit.
From corollary 3.2 of \cite{pattison2023hierarchical}, the SWAPs to perform a single layer of CNOT gates in the syndrome extraction circuit can be done in $\approx 3 \sqrt{n} d$ steps.
As in \cite{pattison2023hierarchical}, we consider a SWAP/idle error rate that is \(1/r\) times the error rate of all other gates.
Accordingly in the first sampling stage, to construct the joint distribution of soft output and logical failures, we sample a surface code memory experiment with \(\tau = \lceil 3 \sqrt{n} d/r \rceil\) rounds of syndrome extraction with \(r=1\) and \(r=10\).

In \cref{fig:hierarchical_surfcode_comparison}, we show a comparison between memories using the hierarchical code, decoded using soft information, and surface codes (simulation details in \cref{app:surface-code-sim}) consuming approximately the same number of physical qubits using phenomenological noise with syndrome measurement error as described in \cref{sec:sim-methods}. 
This corresponds to \num{140} logical qubits encoded in roughly \num{1e5} to \num{2e5} total physical qubits including ancilla qubits in both memories for an overall encoding rate of \(0.7\%\) to \(1.4\%\). 

We find favorable scaling of the hierarchical-code memory logical error rate. If this scaling were to be maintained, we would find that when the hierarchical-code memory achieves a logical error rate of order $10^{-5}$, a surface-code memory of comparable size (140 logical qubits and tens of thousands of physical qubits) has a similar error rate.
However, at very low error rates, the decoder suffers from an error floor; presumably this logical error rate floor for the hierarchical code reflects suboptimal performance of the BP decoding of the outer code in this regime.

Techniques such as ordered-statistics decoding (OSD) post-processing \cite{panteleev2021degenerate}, stabilizer inactivation \cite{du2022stabilizer}, or guided decimation \cite{yao2023belief} have been shown to suppress the error floor for message passing decoders.
Our techniques straightforwardly carry over to this setting.
We caution that these methods have not been proven to \emph{eliminate} the error floor and whether or not a reduced error floor is acceptable depends closely on the precise noise model and application.
Due to compute resource constraints, we would only be able to validate that the error floor is reduced by at most one order of magnitude whereas many practical tasks require many orders of magnitude suppression of the error floor, a regime we are not able to access.

While we analyzed a simplified phenomenological noise model, our qualitative observations should hold in a more realistic noise model such as circuit-level depolarizing noise.
Under circuit-level depolarizing noise, propagation of error during syndrome extraction can cause ``hook errors'' \cite{dennis2002topological} such that the number of resulting errors in the code block is larger than the number of faulty gates in the circuit. Thus hook errors may reduce the total number of faults required to produce a logical error. However, \cite{manes2023distance} has shown that hook errors are not problematic in hypergraph product codes using a naive syndrome extraction circuit with one ancilla qubit per measured check;  in that case, \(d\) circuit faults are required to cover a logical operator.
In our numerics we analyze a generalization of the hypergraph product code family. Hook errors may be similarly benign in this setting.

\section{\label{sec:sampling}%
Error detection using soft output}
Soft-decision decoders are also useful for circuit sampling tasks, as they allow one to reject those samples for which the decoder signals a high likelihood of a logical error. Suppose we desire \(N\) samples drawn from a distribution which is $\epsilon$-close in total variation distance (TVD) to the ideal output distribution of a quantum circuit \(\mathcal{C}\). We execute the circuit fault-tolerantly using a surface code with distance $d$, and assign a soft output to the entire circuit characterizing the probability that any one of the gate gadgets in the circuit has a logical error. Then we
discard samples for which the soft output is below a cutoff value. For fixed $d$, this procedure improves the TVD distance from the ideal distribution. Equivalently, for a fixed target $\epsilon$, the sampling task can be achieved using a smaller value of $d$, reducing the overhead cost of the task.

Up until now we have considered the soft output resulting from decoding a single code block. For the sampling task we desire a soft output pertaining to the entire fault-tolerant circuit. 
To compute a soft output for a large surface code spacetime volume, the soft-decision decoders naturally carry over to the windowed \cite{dennis2002topological} or parallel \cite{skoric2023parallel,tan2023scalable,bombin2023modular} decoder settings: After computing a soft-output for each window, a soft output can be assigned to the entire circuit by taking a union bound.
That is, for soft-outputs \(\{\phi_i\}_i\), let \(q = \sum_i \frac{1}{1+e^{\phi_i}}\) estimate the probability that any one of windows was decoded incorrectly. 
Then, the soft output for the overall circuit is \(\log \frac{1-q}{q}\).

The precise relationship between the numerical value of the soft output and the actual log-likelihood of a decoding failure is expected to depend on the size of the decoding volume. A natural choice is a decoding window with size  \(d\times d \times d\). We therefore envision using sliding-window soft-output decoders acting on windows comparable to this size, with the soft output for the full circuit computed as a function of the soft outputs from all such windows.

\subsection{Repetition code}

To demonstrate this feature, we can leverage the results of \cref{sec:ufd-repcode} to see how a cutoff on $\phi$ affects the logical error rate for the repetition code, assuming a bit flip error rate $p$ and perfect syndrome measurements. From \cref{thm:ufd_repcode}, we know that the soft output is \(\phi = w (n-2 |F|)\), where $F$ is the minimum weight correction and $w=\log\frac{1-p}{p}$ is the edge weight. Therefore, imposing the cutoff $\phi \ge nw\delta$ implies $|F| \le \frac{n}{2}(1-\delta)$, in which case, for a logical error to occur, the number of errors in the code block must satisfy $|E| \ge \frac{n}{2}(1+\delta)$. 
Small values of $\phi$ (and hence of $\delta$) correspond to the regime where the number of errors is close to $\frac{n}{2}$, which dominates the logical error rate when $p$ is small.
Thus by discarding cases where $\phi$ is small, we can suppress the logical error rate substantially. 

Specifically, the joint probability that $\phi$ exceeds the cutoff
($\phi\ge nw\delta$) and a logical error occurs is
\begin{align}
    \Pr\left(|E| \ge \frac{n}{2} (1+\delta)\right)\le 
    \exp\left(-\frac{n}{2}(1 + \delta -2p)^2\right),
\end{align}
using Hoeffding's inequality. Furthermore, as long as $\frac{1-\delta}{2}$ is comfortably above $p$, the event $\phi\le nw\delta$ (i.e. $|E|\ge \frac{n}{2}(1-\delta)$) is exponentially rare for large $n$, so that only a small fraction of runs need to be discarded. Suppose, for example, that the cutoff is chosen so that
\begin{align}
\frac{1-\delta}{2} = p + c,
\end{align}
where $c$ is a positive constant. Then the probability of rejection is upper bounded by $e^{-2nc^2}$, and
our upper bound on the joint probability of acceptance and logical failure becomes
\begin{align}
    \exp\left(-2n(1 -c -2p)^2\right).
\end{align}
For small $c$, this logical error probability in the case with the $\phi$ cutoff imposed is nearly as low as the upper bound that applies in the absence of a cutoff ($\delta=0$) for a block with size $4n$.

\begin{figure}
    \centering
    \includegraphics{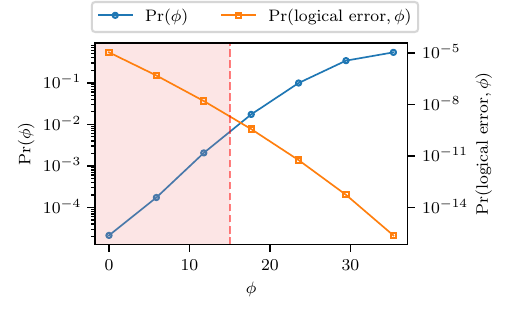}
    \caption{Marginal and joint distributions of \(\phi\) and logical failures under UFD decoding for a length \(n=12\) repetition code with bit flip rate \(p=0.05\) computed using \cref{thm:ufd_repcode}.
    The joint distribution (orange squares, right axis) indicates that the majority of events leading to logical failure are those where \(\phi\) is also small.
    The marginal distribution (blue circles, left axis) indicates that such events are somewhat rare, so discarding them does not cost much.
    The vertical dashed line and red shaded region indicate an arbitrarily selected rejection region containing \SI{0.2}{\percent} of the probability mass which improves the logical error approximately from \num{1e-5} to \num{4e-10}.}
    \label{fig:phi-marginal-rep}
\end{figure}

The improved logical error probability attained by imposing the cutoff on $\phi$ is illustrated in \cref{fig:phi-marginal-rep} for an $n=12$ repetition code and a $5\%$ bit-flip error rate. 
By rejecting about \SI{0.2}{\percent} of the data (\(\phi \le 15\)), the logical error rate is reduced from approximately \num{1e-5} to \num{4e-10}.

Now we consider repeatedly executing a circuit with $V$ logical gates, where a circuit run is rejected if the soft output $\phi$ following any one of the $V$ gates is less than the cutoff value. We can obtain an upper bound on the probability of rejection, as well as an upper bound on the probability of a logical error occurring in a circuit execution that is accepted. For this purpose, we use the tighter Chernoff bound, expressed in terms of the KL divergence.

Define \(D(a||b) = a \log \frac{a}{b} + (1-a) \log \frac{1-a}{1-b}\) to be the KL divergence between two Bernoulli random variables distributed as \(\mathrm{Bernoulli}(a)\) and \(\mathrm{Bernoulli}(b)\), respectively. Our conclusion is expressed in the following theorem. Proof of the theorem and its corollary are deferred to \cref{sec:thm-cutoff-bounds-proof}.

\begin{theorem}\label{thm:cutoff-bounds}
    Let \(\mathcal{C}\) be a quantum circuit with \(V\) gates.
    Consider the circuit \(\mathcal{C}_\mathrm{repp}\) where (1) each qubit has been replaced by a length-\(n\) repetition code that detects bit-flip errors, and  (2) after each gadget corresponding to a gate in \(\mathcal{C}\), stochastic bit-flip noise with rate \(p\) is applied followed by noiseless syndrome measurement and recovery.

    Let \(w = \log\frac{1-p}{p}\).
    Select a relative cutoff \(\delta \in [0, 1-2p)\), and consider the procedure where the output of \(\mathcal{C}_\mathrm{repp}\) is discarded when the soft-output decoder outputs a soft decision \(\phi \le n w \delta\) for any of the \(V\) gates. 
    
    In order to have \(N\) samples at the end of the procedure, \(\mathcal{C}_\mathrm{repp}\) must be sampled \(M\) times where
    \begin{align}
        \mathbb{E}[M] \le\frac{N}{1-V \exp\left[-n D\left(\frac{1-\delta}{2} || p \right)\right]}~.
    \end{align}
    Furthermore, the final measurement outcomes of the $N$ postselected circuits are sampled from 
 a distribution that is at most \(\epsilon\) total-variation distance away from the output distribution of \(\mathcal{C}\) where
    \begin{align}
        \epsilon \le V \frac{\exp\left[-n D\left(\frac{1+\delta}{2}||p\right) \right]}{1-\exp\left[-n D\left(\frac{1-\delta}{2} || p \right)\right]}~.
    \end{align}
\end{theorem}

Without any postselection, the probability of a logical error in each logical gate is bounded above by $\exp\left(-\frac{n}{2}(1-2p)^2\right)$. From the union bound we infer that the probability that a logical error occurs anywhere in a circuit with $V$ logical gates is no larger than $V\exp\left(-\frac{n}{2}(1-2p)^2\right)$. Therefore, to sample from a distribution that is $\epsilon$-close to the ideal distribution, it suffices to choose the length of the repetition code to be 
\begin{align}\label{eq:length-no-postselection}
    n \ge \frac{2\log\frac{V}{\epsilon}}{(1-2p)^2}.
\end{align}

We can reduce the space overhead by using the soft-output signal to reject circuit runs for which the probability of a logical error is unacceptably high. Suppose, for example, that we are willing to discard up to half of all the samples. Then the following corollary of \cref{thm:cutoff-bounds} applies. 

\begin{corollary}
    For the circuit and parameters of \cref{thm:cutoff-bounds}, set
    \begin{align}
        n =  \max
        \left(\frac{2 \log 2V}{(1-2p)^2}, 
        \frac{\left(
        \sqrt{\log\frac{2V}{\epsilon}}+\sqrt{\log 2V} \right)^2}{2(1-2p)^2}\right)\\
    \end{align}
    and
    \begin{align}
        \delta = 1 - 2p - \sqrt{\frac{2 \log 2V}{n}}~.
    \end{align}
    Then, the probability that a sample is discarded is at most \(1/2\) and the remaining samples are drawn from a distribution that is within \(\epsilon\) TVD of the output distribution of \(\mathcal{C}\)
    \label{coro:cutoff-savings}
\end{corollary}

Comparing to \cref{eq:length-no-postselection},
we see that when $\epsilon$ is very small, postselection guided by soft-output decoding reduces the required code length $n$ by nearly a factor of 4. 

\subsection{Surface code}
\begin{figure}[ht]
    \centering
    \includegraphics{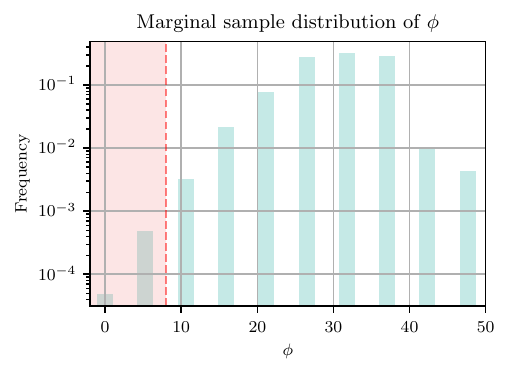}
    \caption{The marginal sample distribution of $\phi$ for a $d=9$ rotated surface code with individual physical qubit error rate $p = 0.005$, faulty measurements (phenomenological noise), $T=9$ syndrome extraction rounds, and $10^7$ samples using UFD.
    Note that the lower tail is exponentially decreasing towards \(0\), as in \cref{fig:phi-marginal-rep}, so the censoring approach in \cref{sec:sampling} will throw out very little data for moderate values of the censoring parameter.
    \label{fig:phi_distr_d9_smallp}}
\end{figure}

We have also investigated the benefits of postselection in surface codes.
\cref{fig:phi_distr_d9_smallp} shows results from simulating a distance-9 rotated surface code memory in the low-error-rate regime using UFD under phenomenological noise with $p=0.005$. Again we observe that small values of $\phi$ are quite rate, where smaller $\phi$ means higher probability of a logical error. Thus we can choose the $\phi$ cutoff to be relatively large, substantially reducing the logical error probability, without needing to discard many samples. 
By discarding the leftmost bins (red), we remove a fraction \(\approx \num{5e-4}\) of the data and achieve an improvement in the logical error rate from $3.0(5) \times 10^{-5}$ to at most $\num{2e-6}$ (\(95\%\) confidence).
Since the tail of the $\phi$ distribution decreases exponentially as $\phi$ approaches 0, as in \cref{fig:phi-marginal-rep}, discarding a larger fraction should improve the logical failure rate much further, but we are unable to resolve such a low logical failure rate in our simulations.

In this example, we are simulating a circuit with only a single gate using a simplified noise model, so these numerics should be interpreted cautiously. But the simulation provides encouraging evidence indicating that postselection guided by the soft output significantly reduces the resources needed to reach a target error rate for the surface code just as for the repetition code.
For surface codes, because the distance grows only as the square root of the block length, the space savings may be more dramatic than for the repetition code.

\section{Discussion and Conclusion}
We have shown how to modify UFD and MWPM decoders for the surface code and repetition code to provide a soft output that estimates the log-likelihood of a decoding failure.
The soft-output algorithm runs in time \(O(V \log V)\) where \(V\) is the number of vertices in the decoding graph.
We have proved that the soft output from UFD is exact for the classical repetition code with bit-flip noise.
For the surface code with MWPM, we have proved that the soft output lower bounds the ratio of probabilities of minimum weight errors in the two equivalence classes.
We supplemented these results with numerics suggesting that the bound is tight up to a multiplicative factor.

We applied this soft-output decoder to the hierarchical code by using the soft information from the inner surface code to aid the decoding of the outer quantum LDPC code.
For a simplified phenomenological noise model, we compared the performance of the hierarchical code, where the outer code is a quasi-cyclic lifted product code, to encoding schemes with the same number of logical qubits using only surface code blocks, finding evidence that the hierarchical code requires fewer physical qubits to reach the same target logical error rate for values of the target error rate of practical interest. 
However, as discussed in \cref{sec:pseudothreshold_sim}), our numerical studies of the hierarchical code encountered a logical error rate floor, presumably due to suboptimal performance in decoding the outer code. Further studies using an improved decoder to compare the performance of the two coding schemes under circuit-level noise  at very low error rates would clarify whether hierarchical codes actually improve the space overhead of fault tolerance in a practical regime.

We also considered applications of the soft-output decoder to circuit sampling tasks, showing that resource requirements can be reduced by discarding circuit runs where the soft information indicates a high probability of a logical error. 
For repetition codes under bit flip noise, we found that the space savings can be as high as \(4 \times\) in the regime where the sampled distribution is required to match the ideal distribution to very high accuracy.
In surface codes, because the distance grows as the square root of the block length, further space savings are expected.
We expect postselection guided by soft information to have broad applications to near-term fault-tolerant quantum computing where space is extremely limited.

\section{Acknowledgements}
We thank Nicolas Delfosse, Michael Newman, and Anirudh Krishna for helpful discussions. 
NM acknowledges funding from Harvard's Herchel Smith Undergraduate Research Program and Caltech's Summer Undergraduate Research Fellowship (SURF) and Information Science and Technology (IST) Venerable WAVE program.
CAP acknowledges funding from the Air Force Office of Scientific Research (AFOSR) FA9550-19-1-0360 and U.S. Department of Energy Office of Science, DE-SC0020290.
JP acknowledges support from the U.S. Department of Energy Office of Science, Office of Advanced Scientific Computing Research (DE-NA0003525, DE-SC0020290), the U.S. Department of Energy, Office of Science, National Quantum Information Science Research Centers, Quantum Systems Accelerator, and the National Science Foundation (PHY-1733907). 
The Institute for Quantum Information and Matter is an NSF Physics Frontiers Center.

\bibliographystyle{apsrev4-2}
\bibliography{bib}

\appendix
\onecolumngrid

\section{Two-stage sampling}
\label{app:sampling}
\begin{figure}[ht]
    \centering
    \includegraphics[width=0.43\textwidth]{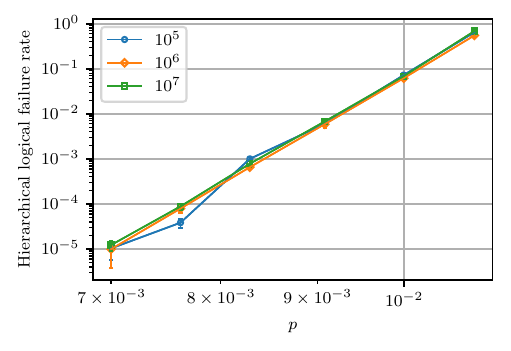}
    \caption{Hierarchical logical failure rates with varying sample size for the empirical $\phi$ distribution. For fixed $r=10$ and distance $d=5$ surface codes, we increase the sample size from $10^5$ to $10^7$ to show that $10^6$ is sufficient for an accurate approximation of the empirical $\phi$ distribution.}
    \label{fig:sampling_phi}
\end{figure}

In this section, we show convergence of the two-stage sampling procedure of \cref{sec:sim-methods} with the number of samples. In \cref{fig:sampling_phi}, we vary the number of samples used to build the $\phi$ distribution. As the number of samples increases, the hierarchical logical failure rates converge to the empirical value. Thus, $10^6$ samples adequately constructs the empirical $\phi$ distribution, and larger sample sizes ($10^7$) do not significantly alter the hierarchical logical failure rates.

\section{Weighted graphs as metric spaces}\label{sec:appendix_graph}
Let \(G = (V, E, \omega)\) be a connected weighted graph such that every weight is positive, i.e. \(\forall e \in E\), \(\omega(e) \ge 0\).
We can consider \(G\) a metric space \((X_G, d)\) by associating to each edge \(e\) the interval \(I_e \equiv [0,\omega(e)] \subseteq \mathbb{R}\) and identifying the endpoints of intervals for which the associated edges are incident to the same vertex.

Concretely, fix an arbitrary orientation of each edge \(o\colon E \to V \times V\).
We define \(X_G = \left(V \sqcup_{e\in E} I_e \right)/ \sim\) where for \((v_1, v_2) = o(e \in E)\), the equivalence relation \(\sim\) is defined such that
\begin{itemize}
    \item \(I_e \ni 0 \sim v \in V \Leftrightarrow v_1 = V\)
    \item \(I_e \ni \omega(e) \sim v \in V \Leftrightarrow v_2 = V\)
\end{itemize}
The metric \(d_G \colon X_G \times X_G \to \mathbb{R}_+\) in this space is then induced by the metric on \(\mathbb{R}\) given by the minimum path between two points.

In the main text we do not distinguish between \(X_G\) and \(G\) nor endpoints of intervals and vertices \(V\).

\section{Quasi-cyclic lifted product codes}
\label{sec:qclp-codes}
The quasi-cyclic lifted product code construction \cite{pantaleev2021quantum} is a generalization of the hypergraph product where for a lift size \(\ell \in\mathbb{N}\), the input matrices are taken to be over the group algebra\footnote{For a finite group \(G\), elements of the group algebra \(\F G\) are given by formal \(\F\)-linear combinations of elements of \(G\).} \(\F\mathbb{Z}_{\ell}\) which is isomorphic to the polynomial ring \(\F[x]/(x^{\ell}-1)\).\footnote{More generally, the matrices are over a group algebra \(\F G\) for some finite abelian group \(G\). Such codes are known as quasi-abelian lifted product codes and are defined in \cite{pantaleev2021quantum}.}

Let \(G\) be a finite abelian group.
For an element of the group algebra \(a \in \F G\), it can be written as \(a = \sum_{g \in \mathbb{Z}_{\ell}} \alpha(g) g\) for some coefficients \(\alpha \colon G \to \F \).
The antipode map is defined to be \(\bar{a}\equiv \sum_{g \in \mathbb{Z}_{\ell}} \alpha(g) g^{-1}\).
For a matrix \(A\) with coefficients in \(\F G\), we define its conjugate transpose \(A^{*}\) to be the antipode map applied elementwise to the transposed matrix \(A^{T}\) i.e. \(A_{ij} = \overline{(A^{*})_{ji}}\).

We also require a means to convert matrices over \(\F\mathbb{Z}_{\ell}\) to matrices over \(\F\).
Let \(W\) be the \(\ell \times \ell\) matrix over \(F\) with entries
\begin{align}W_{ij} = \begin{cases}1 & i+1 \equiv j \mod \ell\\ 0 & \mathrm{otherwise}\end{cases}\end{align}
Then, the map \(\rho \colon \mathbb{Z}_{\ell} \to \F^{\ell\times \ell}\), \(a \mapsto W^{a}\) is a representation of \(\mathbb{Z}_{\ell}\) by \(\ell \times \ell\) circulant matrices over \(\F\).
\(\rho\) induces a map \(\F G \to \F^{\ell\times \ell}\) given by \(\sum_{g \in \mathbb{Z}_{\ell}} \alpha(g) g \mapsto \sum_{g \in \mathbb{Z}_{\ell}} \alpha(g) \rho(g)\).
For a \(n \times m\) matrix \(A\) over \(\F G\), we use the notation \(\rho(A)\) to denote the \(n\ell \times m\ell\) matrix over \(\F\) given by applying the map induced by \(\rho\) elementwise.

For an \(m\) by \(n\) base matrix with entries in \(\F \mathbb{Z}_{\ell}\), the corresponding quasi-cyclic lifted product code is a CSS code defined by the check matrices:
\begin{align}
  H_Z = \begin{pmatrix}\rho(A \otimes I) & \rho(I\otimes A^{*}) \end{pmatrix} & &
  H_X^T = \begin{pmatrix}\rho(I \otimes A^{*})\\\rho(A\otimes I) \end{pmatrix}
\end{align}

We specify the base matrix using the isomorphism \(\F\mathbb{Z}_{\ell} \simeq \F[x]/(x^{\ell}-1) \) induced by \(\mathbb{Z}_{\ell} \ni a \mapsto x^{a}\).
For all numerics, we use a lift size of \(\ell=31\) and a base matrix from reference \cite{tanner2001class}
\begin{align}
  \begin{pmatrix}
    x & x^2 & x^4 & x^8 & x^{16} \\
    x^5 & x^{10} & x^{20} & x^9 & x^{18} \\
    x^{25} & x^{19} & x^7 & x^{14} & x^{28}
  \end{pmatrix}
\end{align}
This code encodes 140 logical qubits into 1054 physical qubits and numerics consistent with a distance of about 20 were observed in \cite{pantaleev2021quantum}.

\section{Simulation of surface codes}
\label{app:surface-code-sim}
Due to limited computational resources, we are unable to directly simulate our baseline (surface codes) at the desired error rates and memory experiment duration, so instead we perform numerical experiments at different parameters and extrapolate.

In order to evaluate the logical failure rate of surface code, we run a memory experiment at a particular physical noise rate and duration.
A memory experiment consists of initializing a quantum memory to a specific logical state, $T$ rounds of syndrome extraction, and then a transversal readout.
A logical failure is recorded if the final state, after correction based on the decoded syndrome information, differs from the initial one.

\begin{figure}[ht]
    \centering
    \includegraphics{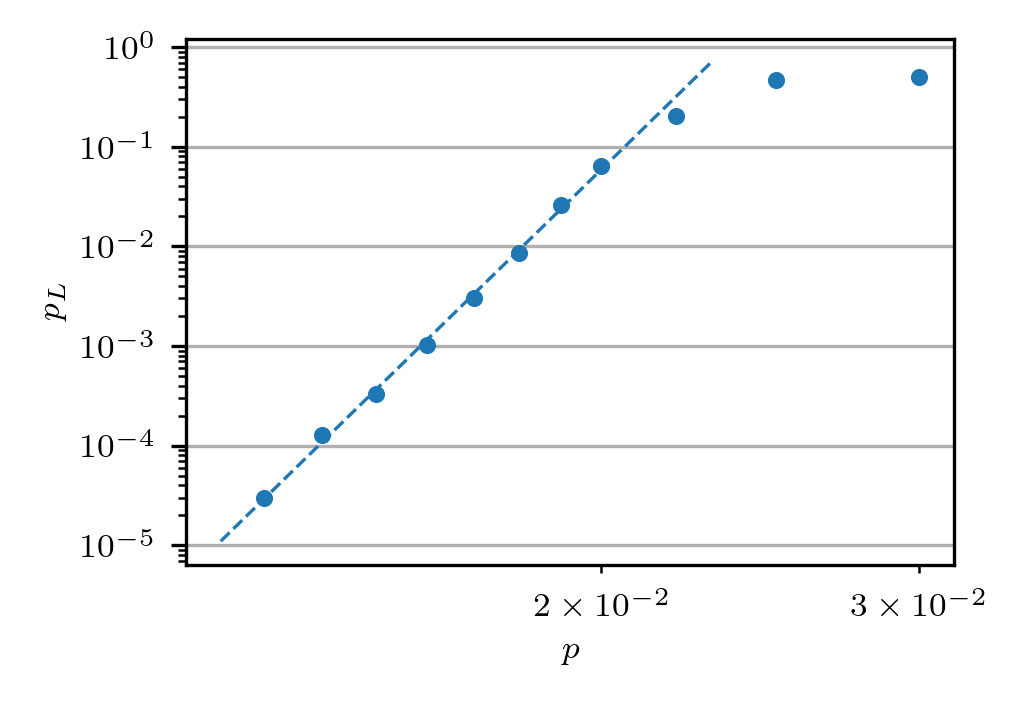}
    \caption{Linear extrapolation of the logical failure rate ($p_L$) as a function of the physical error rate ($p$) for a memory experiment using a distance $d=7$ surface code with $T=33$ syndrome extraction rounds. The linear fit is derived from physical error rates within the interval $0.013 \leq p \leq 0.02$, which is then extrapolated to estimate $p_L$ at low error rates down to $p = 8 \times 10^{-3}$.}
    \label{fig:extrapolate_pL}
\end{figure}

We define the logical failure rate, $p_L$, as the fraction of trials in which logical failures occurred. For small physical error rates $p$, we extrapolate a power-law fit from the below-threshold regime. Figure \ref{fig:extrapolate_pL} illustrates this power-law extrapolation for a surface code of distance $d=7$ and $T=33$ rounds of syndrome extraction.

Linear extrapolation of the logical failure rate ($p_L$) as a function of the physical error rate ($p$) for a memory experiment using a distance $d=7$ surface code with $T=33$ syndrome extraction rounds. The linear fit is derived from physical error rates within the interval $0.013 \leq p \leq 0.02$, which is then extrapolated to estimate $p_L$ at low error rates down to $p = 8 \times 10^{-3}$.

Let $p_L(T)$ denote the probability of logical failure rate for a given number of syndrome extraction rounds $T$.
As the number of error correction rounds increases, the surface code logical failure rate per round approaches a constant for large $T$, specifically $Tp \geq 1$ where $T$ is the number of syndrome extraction rounds, and $p$ is the physical error rate. Thus, for each of the memory experiments, we pick some $T \geq \lceil \frac1p \rceil$.
In the hierarchical code error correction performance comparison (\cref{fig:hierarchical_surfcode_comparison}), we would like to know the failure rate after $T_{\text{mem}}$ syndrome extraction rounds. In the large \(T\) regime, the probability that one of the surface codes failed $p_L(T_{\text{mem}})$ is given by 
\begin{align} p_L(T_{\text{mem}}) \approx \frac{1-\left(1-2p_L(T)\right)^{T_{\text{mem}}/T}}{2}. \end{align}
Next, the logical failure rate of $k$ surface codes is defined to be the probability that at least one of the $k$ surface codes fail, so we plot \begin{align}
    1-(1-p_L(T_{\text{mem}}))^k
\end{align}
as the dashed line in \cref{fig:hierarchical_surfcode_comparison}.

\section{Belief Propagation}
\label{app:belief-propagation}
Belief propagation (BP) is an iterative message-passing algorithm from classical coding theory that is particularly effective in decoding (classical) Low-Density Parity-Check (LDPC) codes (\cite{richardson2008modern} and references therein). It operates on the Tanner graph of an LDPC code. A Tanner graph is a bipartite graph derived from the check matrix of a code. It consists of two types of nodes, variable nodes (corresponding to a bit in the codeword) and check nodes (corresponding to a parity-check equation that the codeword must satisfy).
In a Tanner graph, edges connect variable nodes to check nodes, indicating which bits appear in each parity-check equation.

BP, in the context of decoding LDPC codes, efficiently computes an approximation of the marginals (conditioned on the syndrome) of variable nodes by iteratively passing messages along the edges of the graph between variable nodes and check nodes.
At each step, each node sends a message to its neighbors based on the received messages in the previous step.
The rules to combine received messages at each node are known as the \emph{computation rules}.
We use the product-sum computation rules on binary variables which has the update rules:
\begin{align}
    m^{t+1}_{v_i \to c_j} &:= \ln \left( \frac{1-p}{p} \right) + \sum_{c_{j'} \in \Gamma(v_i) \setminus c_j} m^t _{c_{j'\to v_i}} \\
    m^{t+1}_{c_j \to v_i} &:= (-1)^{s_j} 2 \tanh^{-1} \left( \prod_{v_i' \in \Gamma(c_j) \setminus v_i} \tanh \left( \frac{m^t_{v_i' \to c_j}}{2} \right) \right)
\end{align}
for check node $c_j$ and variable node $v_i$ 
\cite{grospellier2021combining}. $s_i \in \mathbb{F}_2$ is the syndrome result of check node $c_i$, and $\Gamma(c)$ is the neighborhood of the node $c$ in the Tanner graph. 

BP iterations are performed until the bitstring computed from maximizing each marginal matches the syndrome or a predetermined number of iterations have completed.

In the quantum setting, one may operate BP on the alphabet \(\{I,X,Y,Z\}\).
However we are only considering bit flip noise on a CSS code, so we use BP on the binary alphabet \(\{I,X\}\) with the Tanner graph derived from the \(Z\) check matrix.
We refer readers to \cite{richardson2008modern} for a more comprehensive description of BP and its variants, and to \cite{poulin2008iterative} for the application to the stabilizer code setting.

\section{Proof of \cref{thm:cutoff-bounds}}
\label{sec:thm-cutoff-bounds-proof}

\begin{proof}[Proof of \cref{thm:cutoff-bounds}]
    We will use independence of errors and their corrections at different times to apply a union bound over \(V\).

    Consider a single round of errors and error correction with error \(E\), minimum weight correction \(F\), soft output \(\phi\), and cutoff \(n w \delta\) where \(\delta \in [0, 1-2p)\).
    The probability that the sample is discarded is
    \begin{align}
        \Pr(\phi < n w \delta) &= \Pr\left(|F| > \frac{n}{2}(1-\delta)\right)\\
        &\le \Pr\left(|E| > \frac{n}{2}(1-\delta)\right)\\
        &\le \exp\left[-n D\left(\frac{1-\delta}{2} || p \right)\right]
    \end{align}
    Where we have used \cref{thm:ufd_repcode} which implies \(\phi = w (n-2 |F|)\) and a Chernoff bound.

    Furthermore, the probability that there is a logical error given that the sample is not discarded is
    \begin{align}
      \Pr\left(|E| \ge \frac{n}{2} \mid \phi \ge nw \delta\right) &= \frac{\Pr\left(|E| \ge \frac{n}{2} \land \phi \ge nw \delta\right)}{\Pr\left( \phi \ge nw \delta\right)}\\
      &= \frac{\Pr\left(|E| \ge \frac{n}{2} (1+\delta)\right)}{1-\Pr(\phi < n w \delta)}\\
      &\le \frac{\exp\left[-n D\left(\frac{1+\delta}{2}||p\right) \right]}{1-\exp\left[-n D\left(\frac{1-\delta}{2} || p \right)\right]}
    \end{align}
    Where we have again used \cref{thm:ufd_repcode} and a Chernoff bound.

    Putting these results together and using a union bound over the \(V\) events, we conclude that we must execute the circuit 
    \begin{align}
        \frac{N}{\left(1-\exp\left[-n D\left(\frac{1-\delta}{2} || p \right)\right]\right)^V} &\le \frac{N}{1-V \exp\left[-n D\left(\frac{1-\delta}{2} || p \right)\right]}
    \end{align}
    times to achieve \(N\) samples in expectation with better than the cutoff error value.
    These samples are drawn from a distribution that is within
    \begin{align}
      1-\left(1-\frac{\exp\left[-n D\left(\frac{1+\delta}{2}||p\right) \right]}{1-\exp\left[-n D\left(\frac{1-\delta}{2} || p \right)\right]} \right)^V\le
      V \frac{\exp\left[-n D\left(\frac{1+\delta}{2}||p\right) \right]}{1-\exp\left[-n D\left(\frac{1-\delta}{2} || p \right)\right]}
    \end{align}
    total-variation distance from the output distribution of \(\mathcal{C}\).
\end{proof}

\begin{proof}[Proof of \cref{coro:cutoff-savings}]
We begin by loosening the bounds by bounding \(D(p+\epsilon||p) \ge 2 \epsilon^2\) (\cite[lemma~11.6.1]{cover1999elements}) or equivalently replacing the Chernoff bound with the Hoeffding inequality.
    The choice of \(n\) and \(\delta\) ensures that \(\delta \in [0, 1-2p]\):
    \begin{align}
        \delta &= 1 - 2p - \sqrt{\frac{2\log 2V}{n}} \\
        &\ge 1 - 2p - \sqrt{(1-2p)^2}\\
        &\ge 0
    \end{align}

    The probability a sample is discarded is at most
    \begin{align}
        V \exp\left(-nD\left(\frac{1-\delta}{2} || p\right)\right) &\le V \exp\left(-\frac{n}{2} (1 - \delta - 2p)^2\right)\\
        &= V \exp\left(-\frac{n}{2}\left( \frac{2 \log 2 V}{n}\right)\right)\\
        &= \frac{1}{2}
    \end{align}
    The probability that a non-discarded sample contains an error is at most
    \begin{align}
        V \frac{\exp\left[-n D\left(\frac{1+\delta}{2}||p\right) \right]}{1-\exp\left[-n D\left(\frac{1-\delta}{2} || p \right)\right]} &\le V\frac{\exp\left[-\frac{n}{2} \left(2 - 4 p - \sqrt{\frac{2\log 2V}{n}}\right)^2 \right]}{1-\frac{1}{2V}} \\
        &\le 2V \exp\left[-\frac{n}{2} \left(2 - 4 p - \sqrt{\frac{2\log 2V}{n}}\right)^2 \right]
    \end{align}
    Since \(n \ge \frac{2 \log 2V}{(1-2p)^2}\), the quantity inside the parentheses is always positive.
    Define \(\alpha = \frac{\left(
        \sqrt{\log\frac{2V}{\epsilon}}+\sqrt{\log 2V} \right)^2}{2(1-2p)^2}\).
    For \(p \in [0,1/2)\), \(V\ge 1\), the function \(-\frac{x}{2} \left(2 - 4 p - \sqrt{\frac{2\log 2V}{x}}\right)^2 - \log \frac{\epsilon}{2V}\) is monotonically decreasing on \(\left(\frac{2 \log 2V}{(1-2p)^2},\infty\right)\) and has a zero at \(x = \alpha\), so for \(x \ge \alpha\), \(-\frac{x}{2} \left(2 - 4 p - \sqrt{\frac{2\log 2V}{x}}\right)^2 \le \log \frac{\epsilon}{2V}\).
    Returning to the bound and using that \(n \ge \max\left(\alpha, \frac{2 \log 2V}{(1-2p)^2}\right)\), we conclude that 
    \begin{align}
        V \frac{\exp\left[-n D\left(\frac{1+\delta}{2}||p\right) \right]}{1-\exp\left[-n D\left(\frac{1-\delta}{2} || p \right)\right]} &\le 2V \exp\left[\log\frac{\epsilon}{2V}\right]
        \le \epsilon~.
    \end{align}
\end{proof}

\end{document}